\documentclass[aps,a4paper,superscriptaddress,nofootinbib,twocolumn,journal,twoside]{revtex4-1}
\usepackage{verbatim}   
\usepackage{graphicx}
\usepackage{subfigure}
\usepackage{dcolumn}
\usepackage{bm}
\usepackage{booktabs}

\usepackage[colorlinks=true,allcolors=blue]{hyperref}
\usepackage{bm}
\usepackage{amsmath,amsfonts,amssymb}
\usepackage{acronym}
\usepackage{array}
\usepackage{multirow}
\usepackage{soul}
\usepackage{ulem}

\usepackage{comment}
\usepackage{tabularx}
\usepackage{verbatim}
\usepackage{tikz}
\usetikzlibrary{shapes.geometric, arrows}

\tikzstyle{startstop} = [rectangle, rounded corners, minimum width=3cm, minimum height=0.5cm,text centered, draw=black, fill=red!30]
\tikzstyle{process} = [rectangle, minimum width=3em, minimum height=1em, text centered, draw=black, fill=orange!30]

\usepackage{ulem}

\newcommand{\TRC}{
MOE Key Laboratory of TianQin Mission, TianQin Research Center for Gravitational Physics \& School of Physics and Astronomy, Frontiers Science Center for TianQin, Gravitational Wave Research Center of CNSA, Sun Yat-sen University (Zhuhai Campus), Zhuhai 519082, China
}

\definecolor{cerulean}{rgb}{0.0, 0.48, 0.65}

\begin{document}


\title{Detecting Gravitational-waves from Extreme Mass Ratio Inspirals using Convolutional Neural Networks }

\author{Xue-Ting Zhang}
\affiliation{\TRC}

\author{Chris Messenger}
 \affiliation{SUPA, School of Physics and Astronomy, University of Glasgow, Glasgow G12 8QQ, United Kingdom}

\author{Natalia Korsakova}
 \affiliation{ARTEMIS, Observatoire de la C\^{o}te d'Azur, Boulevard de l'Observatoire, 06304 Nice, France}
 
 \author{Man Leong Chan}
 \affiliation{Department of Applied Physics, Fukuoka University, Nanakuma 8-19-1, Fukuoka 814-0180, Japan}

\author{Yi-Ming Hu}
\email{huyiming@mail.sysu.edu.cn}
\affiliation{\TRC}

\author{Jing-dong Zhang}
\affiliation{\TRC}

\date{\today}


\begin{abstract}

Extreme mass ratio inspirals (EMRIs) are among the most interesting gravitational wave (GW) sources for space-borne GW detectors. However, successful GW data analysis remains challenging due to many issues, ranging from the difficulty of modeling accurate waveforms, to the impractically large template bank required by the traditional matched filtering search method. In this work, we introduce a proof-of-principle approach for EMRI detection based on convolutional neural networks (CNNs). We demonstrate the performance with simulated EMRI signals buried in Gaussian noise. We show that over a wide range of physical parameters, the network is effective for EMRI systems with a signal-to-noise ratio larger than 50, and the performance is most strongly related to the signal-to-noise ratio. The method also shows good generalization ability towards different waveform models. Our study reveals the potential applicability of machine learning technology like CNNs towards more realistic EMRI data analysis.

\acrodef{MBH}[MBH]{Massive Black Hole}
\acrodef{sBH}{stellar-mass Black Hole}
\acrodef{MBBH}[MBBH]{Massive Binary Black Hole}
\acrodef{sBBH}{stellar-mass Binary Black Hole}
\acrodef{CO}[CO]{Compact Object}
\acrodef{DWD}[DWD]{Double White Dwarf}
\acrodef{BBH}[BBH]{Binary Black Hole}
\acrodef{BNS}[BNS]{Binary Neutron Star}
\acrodef{BH}[BH]{Black Hole}
\acrodef{NS}[NS]{Neutron Star}
\acrodef{WD}[WD]{White Dwarf}
\acrodef{GW}[GW]{Gravitational Wave}
\acrodef{EMRI}[EMRI]{Extreme Mass Ratio Inspiral}
\acrodef{CNN}{convolutional neural network}
\acrodef{SNR}[SNR]{Signal-to-Noise Ratio}
\acrodef{AGN}{Active Galactic Nuclei}
\acrodef{TDI}{Time Delay Interferometry}

\acrodef{AK}[AK]{Analytic Kludge}
\acrodef{AAK}[AAK]{Augmented Analytic Kludge}
\acrodef{NK}[NK]{Numerical Kludge}

\acrodef{CNN}{Convolutional Neural Network}
\acrodef{FAP}{False Alarm Probability}
\acrodef{TAP}{True Alarm Probability}
\acrodef{ROC}{Receiver Operator Characteristics}

\acrodef{ML}{Machine Learning}
\acrodef{ANN}{Artificial Neural Network}
\acrodef{NN}{Neural Network}
\acrodef{GAN}{Generative Adversarial Networks}
\acrodef{MMA}{Multi-Messenger Astronomy}

\end{abstract}
\maketitle

\section{Introduction}

The successful detection of \acp{GW}  has opened a new window of understanding the Universe \cite{LIGOScientific:2018mvr,Abbott:2020niy,LIGOScientific:2021usb,LIGOScientific:2021djp}.
However, a wide band of \ac{GW} frequencies is still inaccessible for observation.
For example, systems like \acp{DWD} \cite{huang2020science,Korol:2017qcx}, \acp{MBBH} \cite{wang2019science,Klein:2015hvg}, \acp{sBBH} \cite{liu2020science,Sesana:2016ljz}, \acp{EMRI} \cite{Babak:2017tow,Fan:2020zhy},  and stochastic gravitational-wave background \cite{Liang:2021bde,Chen:2018rzo} are expected to emit \ac{GW} signals in the $m$Hz frequency band.
The future detections of these signals can help us better understand the nature of gravity \cite[e.g.][]{shi2019science,Zi:2021pdp} and the Universe \cite[e.g.][]{Zhu:2021aat,Zhu:2021bpp}.
In order to properly understand the physics and astronomy behind \ac{GW} events, 
sophisticated and efficient algorithm to perform data analysis plays a significant role.

Space-borne \ac{GW} missions are capable of detecting \ac{GW} signals in the mHz band \cite{luo2016tianqin,mei2020tianqin}.
Of all expected \ac{GW} sources, \acp{EMRI} are among the most interesting ones.
A typical \ac{EMRI} system is composed of a stellar-mass \ac{CO} and a \ac{MBH}  ($10^4-10^7M_\odot$).
Such systems are believed to be formed in the center of galaxies.
The detection of \acp{EMRI} , among other things,  can help us better understand the growth mechanism of \acp{MBH}, and also put a more stringent constraint on their population properties, like the mass function and the distribution of the surrounding \acp{CO} \cite{Baker:2019nia,AmaroSeoane:2007aw}.  
In addition, the \ac{EMRI} signal can last for thousands to millions of cycles in the $m$Hz frequency band.
This feature empowers the \ac{EMRI} systems to be ideal laboratories to study gravity in a strong regime.
\cite{Zi:2021pdp,Moore:2017lxy,Yunes:2011aa,Canizares:2012is,Barack:2006pq}.

Despite its great scientific potential, there remain great challenges for the end-to-end data analysis of \ac{EMRI} signals.
For example, the ideal \ac{EMRI} waveform should consider the effects of self-force.
Some progress on self-force waveform has been made recently, such as \cite{Lynch:2021ogr} for an eccentric orbit to the second-order, and \cite{Isoyama:2021jjd} for a generic orbit to the adiabatic order. However, a fast and accurate \ac{EMRI} waveform for the most general case is yet to be implemented.
Most of the widely used waveform models are expected to quickly dephase from the physical waveform, which restricts its applicability in \ac{EMRI} data analysis. 
The computational cost of \ac{EMRI} waveforms is also usually very high, 
since it involves orbit integration over a period of months to years.
It has been estimated that in order to implement a classical signal search with
matched filtering, a bank template containing at least $10^{40}$ templates would be needed
\cite{gair2004event}.

Some attempts have been made to explore efficient data analysis algorithms for \acp{EMRI}. Both template-based algorithms and template-free methods have been
proposed to detect the \ac{EMRI} signals. The former includes the  semi-coherent
method \cite{gair2004event} and $\cal{F}{\textendash}$statistic algorithm      
 \cite{wang2012extreme,wang2015first}, and the latter includes the time-frequency algorithm   
 \cite{gair2005detecting, wen2005detecting, gair2007detecting, Gair:2008ec}.
On the parameter estimation side, methods like Metropolis-Hastings search \cite{Gair2008a,babak2009algorithm}, 
parallel tempering MCMC \cite{Ali:2012zz},
and Gaussian process \cite{Chua:2016jnd,Chua:2019wgs} have been implemented.

In recent years, machine learning has gained huge popularity in various complex tasks, such as computer vision \cite{krizhevsky2012imagenet,2014arXiv1409.1556S,2015arXiv151200567S,2015arXiv151203385H,2016arXiv161002357C,2016arXiv160207261S}, speech recognition \cite{deng2013recent}, natural language process \cite{2011arXiv1103.0398C}, and some have been successfully applied to \ac{GW} astronomy \cite{George:2016hay,George:2017pmj,ML_MM_George_2019,Xia:2020vem,Gabbard:2017lja,schafer2020detection,Krastev:2020skk,chan2020detection,Joe2019generalized,Joseph2020robust,Gabbard:2019rde,Williams:2021qyt, George:2017qtr, Shen:2019ohi}. 
Deep learning is a sub-field of machine learning that is based on learning several levels of data representations \cite{Goodfellow_2016_MITpress}, corresponding to a hierarchy of features, factors, or concepts, where higher-level concepts are defined from lower-level ones.
 The \ac{CNN} is a deep learning algorithm, which has been applied in many fields 
 because it can auto\-ma\-ti\-ca\-lly extract data features and easily process various high-dimensional data, such as image classification, target detection, and part-of-speech tagging.  In \ac{GW} data analysis, 
\acp{CNN} has been used to detect \ac{GW} signals detection, such as \ac{BBH} mergers \cite{George:2016hay,George:2017pmj,ML_MM_George_2019,Xia:2020vem,Gabbard:2017lja}, 
\ac{BNS} mergers \cite{schafer2020detection,Krastev:2020skk}, supernova explosions \cite{chan2020detection}, 
and continuous gravitational waves \cite{Joe2019generalized,Joseph2020robust}. 
After being properly trained, deep learning methods also demand low computation cost, which improves efficiency in an end-to-end detection system.
In this work, we use TianQin, a space-borne \ac{GW} detector, as a reference, to explore the \ac{EMRI} signal detection with a \ac{CNN}.

This paper is organized as follows. 
Section \ref{sec:background} introduces the characteristics of \ac{EMRI} and its waveform. 
Section \ref{sec:CNN} provides the architecture of a \ac{CNN}, and discusses how to prepare datasets. 
Section \ref{sec:setup} describes the search procedure. 
The results are presented in Section \ref{sec:Results}. Conclusions and discussions are provided in  Section \ref{sec:conclusion}.

\section{Basics of EMRI detection}\label{sec:background}
\subsection{Basic astronomy of EMRIs}

Galactic nuclei are extreme environments for stellar dynamics \cite{Amaro-Seoane:2012lgq}.
The stellar densities can be as high as $10^6\  \rm M_\odot \ pc^{-3}$ and its velocity scatter can exceed $1000\  \rm km \ s^{-1}$.
Within a galactic nucleus, a large number of \acp{CO}, including \acp{sBH},  \acp{WD}, and \acp{NS}, are expected to exist.
They are considered to be within the ``loss cone" if their orbit would lead to their capture by the central \ac{MBH}, leading to the formation of \acp{EMRI}. 
\acp{CO} outside the loss cone can fill the loss cone through dynamical processes like two-body relaxation  \cite{Amaro-Seoane:2012lgq}.

In addition to this traditional channel, some new formation channels of EMRI systems have also been proposed.
It has been conjectured that a \ac{CO} can be captured by the accretion disk around central \ac{MBH}( i.e., \ac{AGN} disk).
Effects like wind effects, density wave generation, and dynamic friction can assist the inward migration of \acp{CO}, which further boosts the formation of \acp{EMRI} \cite{Pan:2021ksp,Pan:2021oob}.

Although \acp{EMRI} are speculated to exist, no direct observational evidence is available yet (\emph{c.f.}\cite{Arcodia:2021tck}). 
Therefore, currently, the population properties of EMRIs are derived from theoretical modeling.
However, components within the theoretical modeling, like the mass and spin distribution of the central \ac{MBH}, the erosion of the stellar cusp, the M-$\sigma$ relation, etc. are uncertain, which lead to a very uncertain prediction on \ac{EMRI} population \cite{babak2017science}.
Indeed, different assumptions on \ac{EMRI} population lead to orders of
magnitude difference in terms of detection rate by TianQin \cite{Fan:2020zhy}.
Throughout this analysis, an model is adopted to simulate \ac{EMRI} population.
To be exact, we adopt the model M12 from \cite{babak2017science}.

\subsection{Waveform models of EMRIs}
Theoretical modelling of \ac{EMRI} waveforms is essential for the successful detection of \acp{GW}  from these systems.
To accurately compute the \ac{EMRI} waveform, one has to consider the effects of the gravitational self-force, which leads to a deviation of \ac{CO} orbit from the 
test particle geodesics under the Kerr metric of the \ac{MBH} \cite{barack2009gravitational}.
 Unfortunately, waveform simulations with numerical relativity cannot be used for \ac{EMRI} due to prohibitively high computational costs caused by the very high mass ratio of such systems.
 Although some progress has been made in adding self-force into the waveform generation \cite{barack2009gravitational,van2018gravitational,Miller:2020bft,Hughes:2021exa,McCart:2021upc,Lynch:2021ogr,Isoyama:2021jjd}, 
a fully realistic solution for \ac{EMRI} is still yet to be achieved. Nevertheless, there are methods that can generate \ac{EMRI} waveforms. For example, the Teukolsky-based waveform \cite{hughes2005gravitational} and \ac{NK} waveform \cite{babak2007kludge} solves the Teukolsky equation with the geodesic under Kerr metric.
Such methods suffer from low efficiency, while the \ac{AK} method that adopts the post-Newtonian (PN) formula can achieve waveform generation in a much faster way \cite{AK_2004_PRD}.
The \ac{AAK} waveform \cite{chua2015improved} can reach an accuracy similar to \ac{NK} with the generating speed of \ac{AK}.
For our study, since a large number of simulated waveforms are needed for the training, testing, and validation of our \ac{CNN} method, we focus on the analytic family of waveform models and concentrate on both \ac{AK} and \ac{AAK} waveforms.
We remark that during the writing of this manuscript, a new and fast waveform has been implemented \cite{Katz:2021yft}, which could be used to improve efficiency in the future, but we do not expect any significant difference in our findings. 
Besides, \cite{Katz:2021yft} adopts the Schwarzschild \ac{BH} assumption, while we would like to also study the impact of \ac{MBH} spin.

The \ac{AK}  method first calculates the orbit of \ac{CO} through post-Newtonian equations, considering the radiation back-reaction and the Lense-Thirring effect.
The overall evolution of the \ac{CO} orbit is governed by three fundamental frequencies, 
namely the orbital angular frequency $f_{\rm o}$, the periapsis precession frequency $f_{\rm p}$, and the Lense-Thirring precession frequency $f_{\rm LT}$.
The waveform is then calculated based on quadrupolar approximation.
Notice that the overall setup of a waveform is not self-consistent, hence the name of a ``kludge".
However, this choice of formulation allows a fast way of producing waveforms at the cost of its accuracy.

Under the transverse and traceless gauge, two polarizations of \ac{AK}  waveform are defined via a $n$-harmonic waveform which is a function of physical parameters
\footnote{The details of the parameters adopted are presented in Table \ref{tab:AKW-parameters}}:
\begin{align*}
h_+ 
=& \sum_n -\left[ 1+(\hat{L} \cdot \hat{n})^2\right] \left[ a_n \cos2\gamma-b_n \sin2\gamma\right] \\
&+\left[1-(\hat{L}\cdot \hat{n})^2 \right] c_n,\\
h_\times 
=&\sum_n 2(\hat{L}\cdot \hat{n})\left[b_n \cos2\gamma + a_n \sin2\gamma \right]),
\end{align*} 
where  $\hat{L}$ is the orbital angular momentum, $\hat{n}$ is the position of the source, 
and $\gamma $ is an azimuthal angle measuring the direction of pericenter. The coefficients ($a_n, b_n, c_n$) with  $n$ ranges from 1 to 5, 
are determined by the eccentricity  and mean anomaly orbital phase \cite{AK_2004_PRD}. The orbital phase is determined by three fundamental frequencies $f_{\rm o},\ f_{\rm p},$ and  $f_{\rm LT}$.  
More details on waveform generation can be found in \cite{AK_2004_PRD, Fan:2020zhy}. 
Notice that since the self-force can not be properly incorporated by the \ac{AK} waveform, the spin parameter for the \ac{CO} is not included.

The \ac{AK} waveform quickly deviates from physical systems, due to intrinsic inconsistency.
Under the \ac{AK} model, instead of using the physical parameters, one can adopt the effective parameters to evolve the orbit and produce a waveform similar to the \ac{NK} model. 
The \ac{NK} model is more accurate as it combines PN orbital evolution with Kerr geodesics.
Chua et al. implemented the \ac{AAK} waveform by establishing the link between the physical parameter and the effective parameter.
To obtain such a link, one can evolve the \ac{NK} waveform for a short period and determine the mapping relationship to derive the correct fundamental frequencies $f_{\rm o}$, $f_{\rm p}$, and $f_{\rm LT}$ \cite{Chua:2015mua}.
In this way, the \ac{AAK} waveform can achieve both the accuracy of \ac{NK} and the speed of \ac{AK}.

\subsection{The TianQin mission}
For a given gravitational wave detector, one can use the antenna pattern to describe its response to a source at a given location, polarization, and frequency \cite{cutler1998angular,cornish2001space,cornish2003lisa}.
A typical \ac{EMRI} signal lasts for a long time.
If we observe from the point of view of the detector, then a given source would change in location as well as in polarisation.
One needs to know the orbit of the spacecraft to derive the overall evolution of the signal as seen at the detector.
In this work, we focus our attention on the proposed TianQin mission \cite{luo2016tianqin,zhang2020full} for the following analysis.

The TianQin mission is designed to be a constellation with three satellites, which accommodate test masses in free fall and use drag-free control to follow these test masses.
The satellites share an Earth-central orbit and form a regular triangle. They are connected by laser links and can form Michelson interferometers to detect \ac{GW}. 
The geocentric nature of the TianQin orbit makes it adopts a ``3-month on 3-month off" work scheme, that the detectors would operate continuously for three months followed by three months of break.
Each pair of neighboring laser arms can form interference on the vertex satellite, whose noise can be described by the power spectral density (PSD):

\begin{equation}\label{eq:noise_curve}
\begin{split}
S_n(f) =& \frac{1}{L_{\rm arm}^2}\left[ \frac{4S_a}{{(2\pi f)^4}} \left(\frac{1+10^{-4}\  \text{Hz}}{f}\right) + S_x \right] \\ & \left[ 1+ 0.6\left(\frac{f}{f_*}\right)^2\right],
\end{split}
\end{equation}
where $f$ is the frequency, $S_a=1\times 10^{-30}\ {\rm m^2 \  s^{-4}/
\text{Hz}}$ describes the acceleration noise, $S_x=1 \times 10^{-24}\ \rm m^2/\rm Hz$ describes the positional noise, and $f_*=\frac{c}{2\pi L_{\rm arm}}$ is the transfer frequency with $c$ being the speed of light and $L_{\rm arm}=\sqrt{3}\times 10^8\  {\rm m}$ is the arm length.

For the low frequency, where $f<f_*$, the three interferometers can be combined into two orthogonal channels, corresponding to the two GW polarisations.
So the recorded GW signal can be expressed as:
\begin{equation}\label{eq:hI}
\begin{split}
  h_{I,II}(t)=\frac{\sqrt{3}}{2}\Big(h_+(t)F_{I,II}^+(\theta_S,\phi_S,\psi_S)\\
    +h_{\times}(t)F_{I,II}^{\times}(\theta_S,\phi_S,\psi_S)\Big),
\end{split}
\end{equation}
where $F_{I,II}^+$ and $F_{I,II}^{\times}$ are the antenna pattern for both
plus and cross polarisations, and
\begin{equation}
\begin{split}
F_I^+(\theta_S,\phi_S,\psi_S) =&  \frac{1}{2}(1+\cos^2 \theta_S)(\cos2\phi_S \cos2\psi_S \\
& - \cos \theta_S \sin2 \phi_S \sin 2\psi_S). \\
F_I^{\times}(\theta_S,\phi_S,\psi_S) =& \frac{1}{2}(1+\cos^2 \theta_S)(\cos2\phi_S \sin 2\psi_S \\
&+ \cos\theta_S \sin2\phi_S \cos2\psi_S).
\end{split}
\end{equation}
Here $\theta_S$ and $\phi_S$ are the sky location of the source, and $\psi_S$ is the polarization angle.
All quantities are defined in the detector frame, which makes it
time-dependent \cite{liu2020science}.

The two effective channels differ only in polarisation angles by $\pi/4$, which means
\begin{equation}
\begin{split}
    F_{II}^{+}(\theta_S,\phi_S,\psi_S) = F_I^+(\theta_S,\phi_S,\psi_S-\pi/4), \\
    F_{II}^{\times}(\theta_S,\phi_S,\psi_S) = F_I^{\times}(\theta_S,\phi_S,\psi_S-\pi/4).
\end{split}
\end{equation}

Finally, a correction term is added to account for the Doppler effect induced
by the orbital motion of the detectors \cite{liu2020science}.
\begin{equation}
\Phi_{\rm Doppler} = 2 \pi f(t) \times R \times \sin{\bar{\theta}_S}  \times \cos{\left( \bar{\Phi}(t) -\bar{\phi}_S \right)}, \\
\end{equation}
where $f$ is the frequency of \ac{GW}, $R = 1 $AU and barred symbols denote parameters in Solar System Barycentre frame, with $\bar{\Phi}(t) = \bar{\phi}_0 +2\pi t/T,\  T= 1 \rm year $ is Earth's orbital period around the Sun, and  $\bar{\phi}_0 $ is the initial location of TianQin at time t = 0. Notice that in higher frequencies when $f\ge f_*$, the long-wavelength approximation breaks
down and the two channels are no longer uncorrelated, hence an extra correction term is
needed in Equation \ref{eq:hI} \cite{cornish2001space,cornish2003lisa}.  
Fortunately, \ac{EMRI} signals we are interested in, are located mostly within the low-frequency range.
This is because the highest orbital frequency of CO is determined by the last stable orbit(LSO), where $\nu_{\rm LSO}= 1/(2 \pi M) \times 6^{3/2}$.
Given a typical \ac{MBH} with $10^5 M_{\odot}$, $\nu_{LSO}=0.04 <f_*=0.28$ Hz.
Therefore, we stick
with the long-wavelength approximation throughout the work.

\begin{figure}[!htbp]
\centering
\includegraphics[scale=0.45]{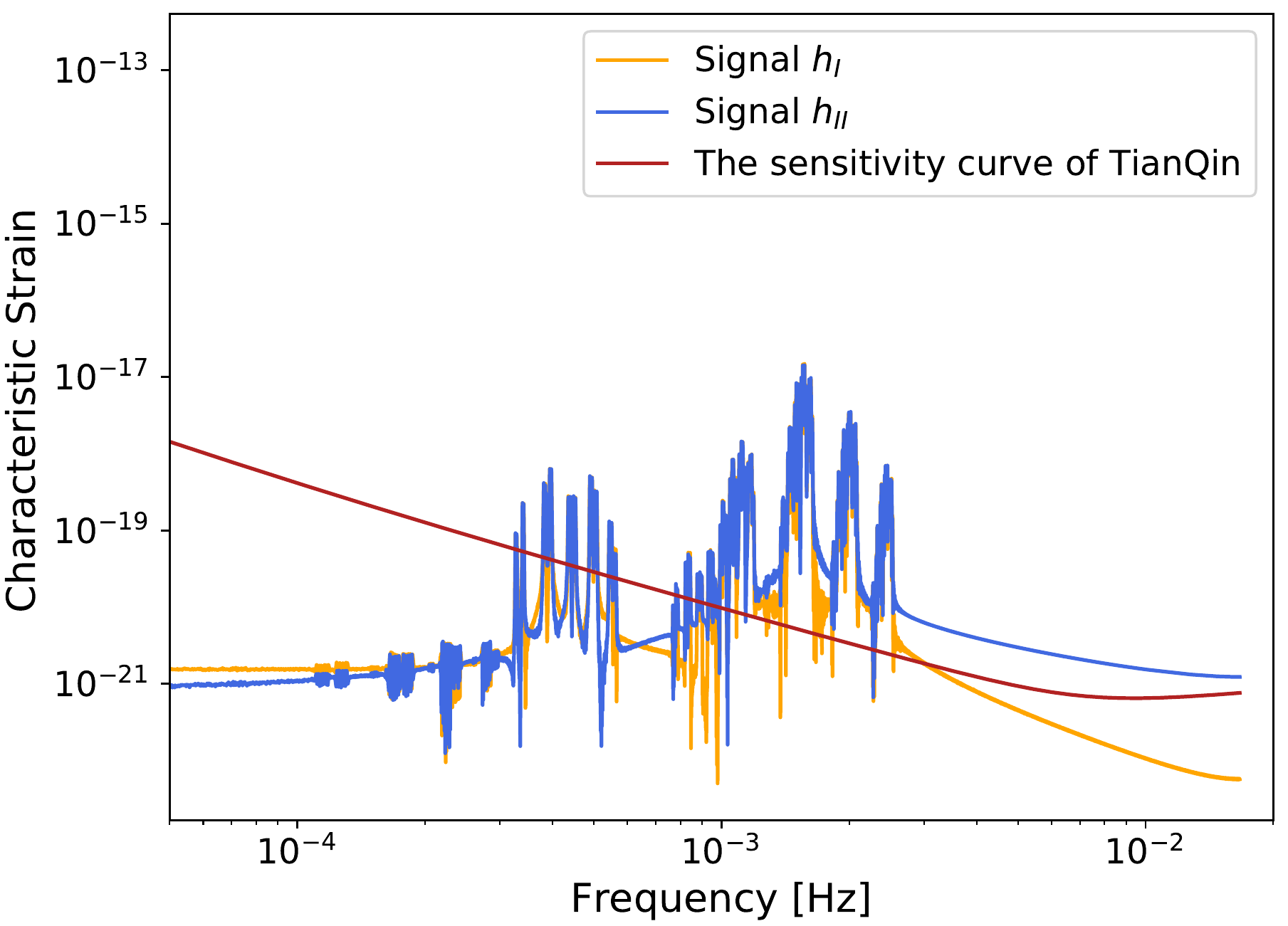}
  \caption{\label{fig:signals-fd} An example \ac{EMRI} signals compared with the sensitivity curve of TianQin. A total length of 3 months observation time is assumed.  }
\end{figure}

As an illustration, we show in Figure \ref{fig:signals-fd} an example \ac{EMRI} signal (on both channels) on the TianQin sensitivity curve. 
This \ac{EMRI} system consists of a 10 $M_{\odot}$ \ac{BH} and \ac{MBH} with 4.5 $\times 10^6\  M_{\odot}$ and a spin of 0.97. The \ac{SNR} of this signal equals 50.

We define the overall \ac{SNR} as the root sum square of inner products of waveform with itself across two channels of TianQin,
\begin{equation}
\begin{split}
\label{Optimalsnr}
  \rho &=\sqrt{ \sum_{i={\rm I,II}} \langle {h}_i |  {h}_i\rangle} \\
  &=\sqrt{ 4 \Re \sum_{i={\rm I,II}} \int^{\infty}_{0}{\frac{\tilde{h}_i(f)\tilde{h}_i^*(f)}{S_n(f)}}{\rm d}f} \\
  &\approx \sqrt{ 4 \sum_{i={\rm I,II}}  \int^{f_{\rm high}}_{f_{\rm low}}\left(\frac{\tilde{h}_i(f)}{\sqrt{S_n(f)}}\right)\left( \frac{ \tilde{h}_i(f)}{\sqrt{S_n(f)}} \right)^*{\rm d}f }.
\end{split}
\end{equation}

Here we use `` $\langle g|h \rangle$" to represent the inner product between the two frequency domain waveforms $\tilde{g}(f)$ and $\tilde{h}(f)$.
Notice that in the last line, the \ac{SNR}  in a channel is simply the integration of squared modulus of whitened signal over the frequency range between $f_{\rm low}=10^{-4}\ \rm Hz$  and $f_{\rm high}=1\ \rm Hz$, which are sensitivity frequency band of TianQin.

\section{convolutional neural networks for detection}\label{sec:CNN}

\subsection{Data preparation }\label{sec:data_prep}
For the detection of the signal in noise we are going to use \acp{CNN}.
In order to train and verify \ac{CNN} properly, one needs to divide the data into three groups, namely {\it training data}, {\it validation data}, and {\it testing data}.
In these datasets, half of them are signal+noise samples, and the rest are noise-only samples. 
Using training data, the \ac{CNN} learns how to classify samples. 
Validation data is incorporated in the
training process to verify that the \ac{CNN} is learning correctly. Finally,
testing data can test the efficiency of the trained \ac{CNN}.

\begin{figure*}[!htbp]
\centering
\includegraphics[scale=0.65]{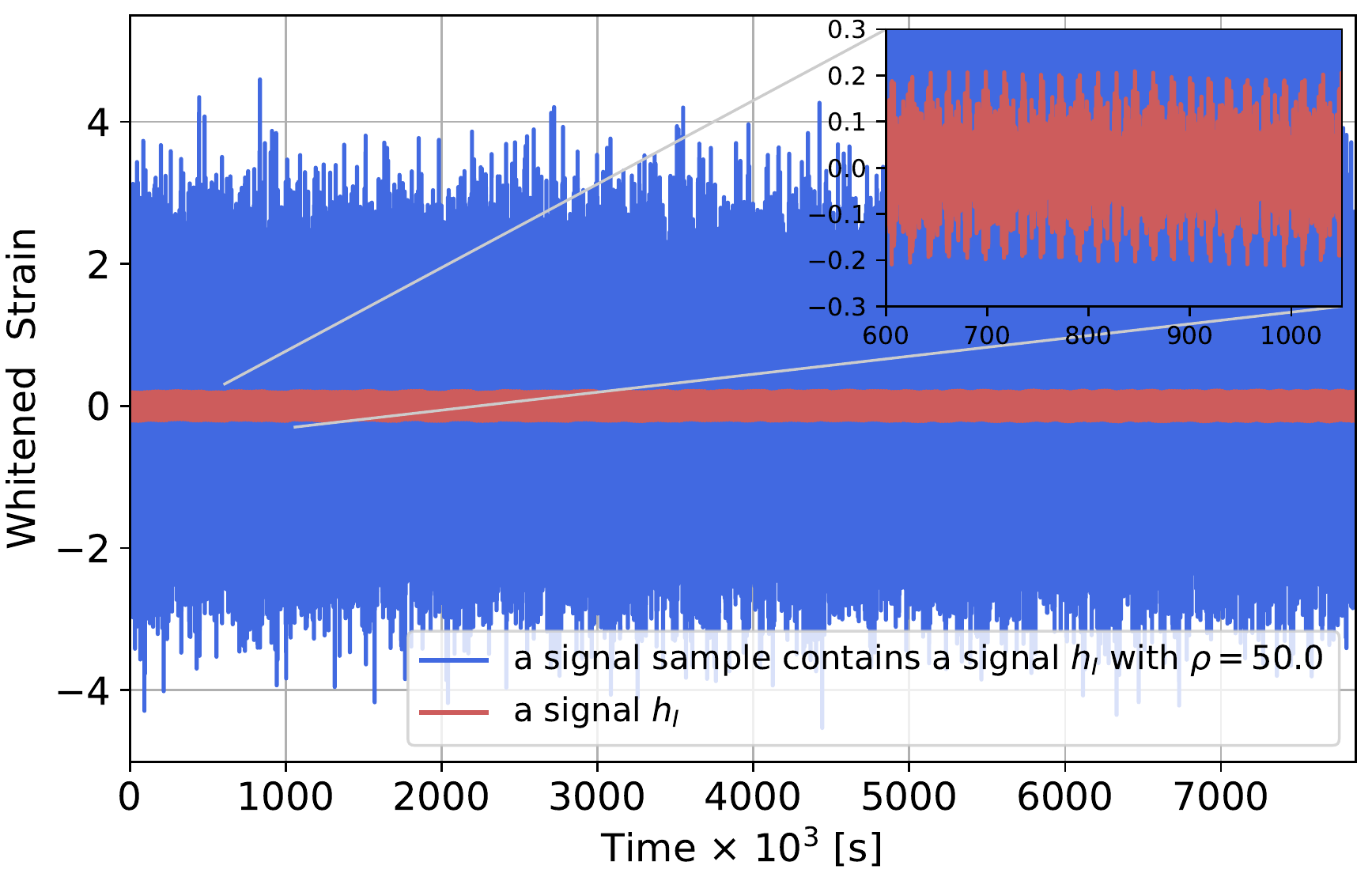}
  \caption{\label{fig:inupt-data} An example of whitened data in channel $I$ in comparison with signal $h_I$ alone.
  For this event, the \ac{SNR} is set to be 50.
  We draw the reader's attention to the difference in scale for the noise and
the signal.}
\end{figure*}

In our case, we aim to accomplish the task of classifying the data into two categories: with signal, or without signal.
One can express the data $d$ as the addition of random Gaussian noise $n$ and the GW signal $h$ (in the case it is present in the data):
\begin{equation}\label{eq:data}
d(t) =\left\{
\begin{aligned}
  & h(t) + n(t) ,  \text{ if signal is present}\\
  & n(t),  \text{ if there is no signal}.\\
\end{aligned} 
\right.
\end{equation}
The raw output of TianQin is three correlated time domain data channels.
After some preprocessing they can be reduced to two channels with uncorrelated noise, plus a channel with a response to astrophysical signals highly suppressed.
This process is implemented by the \ac{TDI} \cite{PhysRevD.62.042002,PhysRevD.59.102003}. 
Out of simplicity, we do not implement \ac{TDI}, but rather treat TianQin as two orthogonal Michelson interferometers.
In Figure \ref{fig:inupt-data}, we illustrate a data sample that contains an \ac{EMRI} signal, using the parameters demonstrated in Figure \ref{fig:signals-fd}.

We prepare the generation of noise and signal separately. Due to the ``3-month on 3-month off" work scheme of TianQin, the longest chunk of continuous data lasts only three months. To avoid the complexity introduced by considering the data gap issue, we focus our analysis on data length of three months (or 7864320 seconds). 
By adopting the long-wavelength approximation, 
it is also sufficient to assume the sample rate of $1/30$ Hz, which leads to the data size of 262144 samples.
The parameters are summarised in Table \ref{tab:data-setting}.

\begin{table}[ht]
\centering
\caption{\label{tab:data-setting} The setting of data simulation.}
\begin{ruledtabular}
\begin{tabular}
{p{0.06\textwidth}<{\centering}%
p{0.2\textwidth}<{\centering}%
p{0.2\textwidth}<{\centering}}
\textbf{Symbol} & \textbf{ Physical Meaning} & \textbf{ Range}  \\
\hline  
$T$     &  observation time  & 7864320 seconds \\ 
$f_s$	&	sample rate 	& 1/30 Hz \\
$N_t$	&  a data size 	& 262144
\end{tabular}
\end{ruledtabular}
\end{table}



\begin{table*}[ht]
\centering
\caption{\label{tab:AKW-parameters} Summary of physical parameters and their meaning under the analytic kludge model.}
\begin{ruledtabular}
\begin{tabular}
{p{0.06\textwidth}<{\centering}%
p{0.46\textwidth}<{\centering}%
p{0.42\textwidth}<{\centering}}
        \textbf{Symbol} & \textbf{Physical Meaning} & \textbf{ distribution}  \\
\hline  $M$     &  \ac{MBH} mass & uniform in $\ln{M}$ over [$10^4\ M_{\odot}$,$10^7 \ M_{\odot}$]\\ 
        $\mu$   & \ac{CO} mass  & $10\ M_{\odot}$ \\ 
        $a$ & \ac{MBH} spin magnitude, defined as $|\vec{S}|/M^2$, with $\vec{S}$ being the \ac{MBH} spin angular momentum & uniform [0,1] \\ 
        $e_{\rm lso}$  & orbital eccentricity at plunge & uniform [0,0.2]\\ 
        $\Phi_0$  & mean anomaly at plunge & uniform[$0,2\pi$]\\ 
        $\alpha_0$  & azimuth angle for orbital angular momentum $\vec{L}$ at plunge & uniform[$0,2\pi$]\\ 
        $\lambda$  & angle between $\vec{L}$ and $\vec{S}$ & cos($\lambda$)= uniform[-1,1]\\ 
        $\gamma_0$ & angle between $\vec{L}\times \vec{S}$ and pericenter at plunge & uniform[$0,2\pi$]\\ 
        $\nu_{\rm lso}$  &  orbital frequency of the last stable orbit & depend on $e_{\rm lso}$\\ 
        $\theta_S$  & source direction's polar angle & uniform[$0,2\pi$]\\ 
        $\phi_S$  & azimuthal direction to source & cos($\phi_S$)=uniform[$-1,1$]\\ 
        $\theta_K$  &  the polar angel of \ac{MBH}'s spin  & uniform[$0,2\pi$]\\ 
        $\phi_K$  & azimuthal direction of \ac{MBH}'s spin & cos($\phi_K$)=uniform[$-1,1$]\\ 
        $t_c$ & merger time of a \ac{CO} plunges into \ac{MBH} & fixed to 7864320 seconds \\
        $D_L$  &   luminosity distance to source & irrelavent as will be rescaled to \ac{SNR} with uniform[50,120]\\ 
\end{tabular}
\end{ruledtabular}
\end{table*}

The detector noise was assumed to be Gaussian and was generated by sampling from the one-sided PSD given by Equation \ref{eq:noise_curve}.
For the simulation of an \ac{EMRI} waveform signal, we used both \ac{AK} and \ac{AAK} models.
For different purposes that we will detail in the following sections, we generate 7 groups of signals, whose properties are explained in Table \ref{tab:simulated_wave}, and the shared default parameter distributions are listed in Table \ref{tab:AKW-parameters}.
The data are whitened in the frequency domain. When data are converted back to the time domain, we apply a Tukey window function with the coefficient $\alpha =1/4$ to avoid the spectrum leakage.  

Notice that Table \ref{tab:AKW-parameters} does not reflect an
astrophysical model, and one might argue that an astrophysically motivated distribution for \ac{EMRI} parameters could be more realistic. 
However, an astrophysically motivated distribution is not necessarily the best
choice for training the \ac{CNN}. The reason for that is that the astrophysically motivated models are subject to huge uncertainty. Therefore sticking with any specific model runs a risk of biasing the network.

\ac{SNR} of a signal is inversely proportional to the distance.
 When assessing the performance of a \ac{CNN}, it is convenient to compare signals with similar \acp{SNR} than with similar distances.
Therefore, we do not fix distribution of distance $D_L$ in Table \ref{tab:AKW-parameters}. 
The choice of \ac{SNR} range is determined by balancing the hardware requirement between data storage, waveform generation speed, and I/O speed in training.

\subsection{Mathematical model}

\ac{CNN} is a specific realization of \ac{NN} which features the
inclusion of convolution operations~\cite{ng2013unsupervised,Goodfellow_2016_MITpress}. 
In \ac{NN}, each ``neuron" represents a non-linear mapping or an affine transformation function.
The non-linearity is achieved by an activation function.
A typical realization of a \ac{CNN} includes convolution layers, pooling layers, and fully-connected layers (or dense layers) \cite{lecun2015lenet}. It can be treated as a classifier. 
Such a \ac{CNN} composed of multiple neurons can be understood as a non-linear function  $f_w(\cdot)$ that maps the input space of the data to the output space:
\begin{equation}
y_i = f_w(d_i),
\end{equation}
where $y_i$ is the output probability for the  $i$-th data $d_i$ to belong to a certain class, $w$ are weights of the CNN.

Based on training data a \ac{CNN} can learn features of the input data by updating the weights associated with
different neurons by minimizing the difference between the output prediction and training data labels.
The size of our training data is large, therefore we cannot optimize weights with all the data at once and need to adopt mini-batching.
The training process is divided into multiple epochs. 
For each epoch, the weights are  updated by the optimizer algorithm. 
When the training process converges, the \ac{CNN} can classify new data into different categories.

To achieve the reliable ability to classify, one needs, first, to define the loss function $L(y_{\rm true}, y_{\rm pred})$, which quantifies the difference between the \ac{CNN} output and labels in the training process.
This could be done by adopting the binary cross-entropy function:
\begin{equation}\label{eq:loss_function}
    L = \sum_{j=1}^{N} \left [-y^j_{\rm true} \log(y^j_{\rm pred}) -(1- y^j_{\rm true}) \log(1-y^j_{\rm pred}) \right ],
\end{equation}
where $N$ is the total number of training data samples,
for the  $j$-th data instance. The noise-only samples are labeled as $y^j_{\rm true}=0$, and samples containing signals are labeled as $y^j_{\rm true}=1$.
The output of \ac{CNN} $y^j_{\rm pred}$ represents the assigned probability for each category.
The loss function is minimized when the predicted value for the label class matches the one from the training data with the highest probability.
In the training process, the weights are updated through the {\it Nadam} algorithm, such that the loss function can be minimized by following the adaptive gradient descent with Nesterov momentum \cite{dozat2016incorporating}. 
During training, validation data is used to test the trained \ac{CNN} after each epoch and use those result to evaluate the performance of the \ac{CNN} 
and its stability towards overfitting.

\subsection{The \ac{CNN} architecture}
In a \ac{CNN} different layers play different roles. 
In the {\it convolution layer}, features can be extracted by applying
convolution between the input and the convolution kernel.
After multiple convolution layers with various filters, a \ac{CNN} can learn different features of the input data from lower-level to higher-level.
The purpose of the {\it pooling layer} is to reduce the size of the convolved features while remaining the key features so that one can reduce computation complexity and avoid over-fitting.
In the {\it fully-connected layer}, neurons are connected with all activation values to the previous layer, with the final output indicating the predicted class probability.
In this work, {\it rectified linear unit (relu)} is selected as the activation function to all but the output layer, for which the {\it softmax} function is adopted. 
Relu function is convenient and efficient to avoid vanishing gradient problems when training a \ac{CNN} \cite{krizhevsky2012imagenet}, and softmax function at the last layer achieves a normalization so that the output is constrained in the range of (0,1). 
In this way, the output can appropriately represent 
the probability that the input sample belongs to one of the classes. The performance of a \ac{CNN} depends on many hyper-parameters, like \ac{CNN} depth, convolutional layer number, number of kernels, and their respective size.

\section{Search procedure}\label{sec:setup}

In this section, we will explain the details about the \ac{CNN} application to
the \ac{EMRI} signal search
\footnote{In our implementation of the \ac{CNN}, we use Keras 2.2.4, and Tensorflow 1.1.18. 
We implement our software on top of a  
GPU Tesla V100 PCIe 16 GB, CPU Intel Xeon Gold 6140 (72) @ 2.3, with a memory of 256GB.}.

\subsection{Training phase}

In order for \ac{CNN} to gain the classification ability, a certain amount of input data is necessary for the training.
More samples are needed to train \ac{CNN} to detect weaker signals. 
In the actual training, it turns out that the I/O of training data becomes the bottleneck.
Therefore, we are limited in the size of total data, which in turn determines the \ac{SNR} lower limit.
After some trial and error, we converged  at 50 as the \ac{SNR} lower limit in our training sample.
On the other hand, stronger signals can be detected more easily.
We concentrate on signals with \ac{SNR} comparable with the lower limit, therefore an upper limit of \ac{SNR} 120 is adopted.
The total amount of $2.5 \times 10^5$ {\it signal plus noise} samples are used to train the final \ac{CNN}.

With the data ready, we need to train the network to maximize the classification ability without over-fitting.
In general, a more complicated network is expected to perform better on the training data.  However, a too-complicated neural network runs a risk of over-fitting or being too acquainted with the training data so that the
performance on training data is not generalizable towards unfamiliar data.
Therefore one needs to check the consistency of the \ac{CNN} performance on both training data and validation data. 
One can identify the over-fitting if the performance on the training data is significantly better than the validation data. 
This can be relieved by adopting a less complicated neural network, adding dropout layers, or increasing the size of the training set.

Finally, we need to fix the training architecture.
We start our trial of convolution kernel size of (1,5) as suggested by previous studies 
\cite{Gabbard:2017lja,Joseph2020robust}, and finally settled with a size of (1,34) through trial and error.
We apply the max-pooling in the pooling layer, which can retain the maximum value of each patch of the input.
The mini-batch size is set to 56, and we train the \ac{CNN} with 300 epochs.  
For each mini-batch, 56 simulated datasets will be randomly selected from samples, half of which are {\it signal plus noise} samples and the
other half are {\it noise-only} samples.  The wall clock time for each epoch of
training is about 2 hours.  Strategies like early stopping \cite{yao2007} are also employed to
increase efficiency.  In our experiments, the training process will be stopped
when either 300 epochs are reached, or the accuracy of validation data no
longer increases for 50 epochs.
Finally, it takes about 10.5 days to train the final \ac{CNN}. This trained model can afterwards be used for testing.
We summarise the architecture of the \ac{CNN} in Table \ref{tab:CNN_architecture}.

\begin{table}[ht]\footnotesize
\caption{\label{tab:CNN_architecture}the architecture of the \ac{CNN}}
\centering
\begin{ruledtabular}
\begin{tabular}{lcp{0.08\textwidth}<{\centering}
cp{0.08\textwidth}<{\centering}}
        & \textbf{ Layers} & \textbf{kernel number}  &\textbf{kernel size} &\textbf{Activation function} \\
\hline
        1 & Input       &/  &  matrix(size: $2 \times 262144$ ) & / \\   
        2 & Convolution &32 & matrix(size:$1 \times 34$) & \text{relu}\\ 
        3 & Pooling    &16 & matrix(size:$1  \times 8$) & \text{relu} \\ 
        4 & Convolution &16 & matrix(size: $1  \times 8$) & \text{relu} \\ 
        5 & Pooling    &16 & matrix(size: $1 \times 6$)  & \text{relu} \\ 
        6 & Convolution &16 & matrix(size: $1 \times 6$) & \text{relu}\\ 
        7 & Pooling   &16 & matrix(size: $1 \times 4$)  & \text{relu}\\ 
        8 & Flatten &/ & / & / \\ 
        9 & Dense &/ & vector(size: 128)  & \text{relu}\\ 
        10  & Dense &/ & vector(size: 32) & \text{relu}\\ 
        11 &  Output &/ & vector(size: 2) & \text{softmax} \\
\end{tabular}
\end{ruledtabular}
\end{table}

\begin{table*}[!htbp]
\caption{
 Different waveform setups used as testing data. 
 The first column indicates the index; the second column is the waveform model; 
 the third column describes the parameter distribution, where to some astrophysical parameters the range shown in the table will be applied, but other unspecified parameters will remain in the same range as the training data( See table \ref{tab:AKW-parameters}). 
}\label{tab:simulated_wave}
\begin{ruledtabular}
\begin{center}
\begin{tabular}
{p{0.07\textwidth}<{\centering}%
p{0.09\textwidth}<{\centering}
p{0.1\textwidth}<{\centering}
p{0.1\textwidth}<{\centering}
p{0.1\textwidth}<{\centering}
p{0.1\textwidth}<{\centering}
p{0.16\textwidth}<{\centering}}
{\centering \textbf{number}} & 
{\centering \textbf{waveform model}} &
\multicolumn{4}{c}{\centering \textbf{physical parameters distribution}} & 
{\centering \textbf{signal samples number}} \\
\cline{3-6}
\hline
1  & \ac{AK} & \multicolumn{4}{c}{\centering $\rho \in$ uniform [50,120]} &  500  \\
2  & \ac{AK} & \multicolumn{4}{c}{\centering  $\rho$ $>50$, astrophysical model M12} & 500  \\
3  & \ac{AAK}& \multicolumn{4}{c}{\centering  $\rho \in$ uniform [50,120]} & 500  \\
4  & \ac{AK} &\multicolumn{4}{c}{\centering $\rho$ enumerates $10, 20, ..., 130$} & 1000 $\times 13$  \\
\multirow{3}*{\centering 5}  &\multirow{3}*{\centering \ac{AK}} &\multicolumn{3}{c}{\multirow{3}*{\centering  $M$ enumerates $10^4, 10^{4.5}, ...,  10^7M_{\odot}$, $a = 0.98$}} & $z =$ 0.1  & \multirow{3}*{1000 $\times 7 \times 3$ }\\
~  & ~ &\multicolumn{3}{c}{~}& $z =$ 0.2 &~ \\
~  & ~ &\multicolumn{3}{c}{~}&  $z =$ 0.3 &~\\
\multirow{3}*{\centering 6}  &\multirow{3}*{\centering \ac{AK}} &\multicolumn{3}{c}{\multirow{3}*{\centering  $M = 10^6M_{\odot}$, $a$ enumerates $ 0.0, 0.2, 0.4, 0.6, 0.8, 0.98$}} &\centering $z =$ 0.1 & \multirow{3}*{1000 $\times 6 \times 3$ }\\
~  & ~ &\multicolumn{3}{c}{~}& $z =$ 0.2  &~ \\
~  & ~ &\multicolumn{3}{c}{~}& $z =$ 0.3  &~\\
\multirow{3}*{\centering 7}  &
\multirow{3}*{\centering \ac{AK}} &
\multicolumn{4}{c}{\centering  $M = 10^{5.5}M_{\odot}, a = 0.98$, $z$ enumerates $0.1, 0.2, 0.3$}  & \multirow{3}*{1000 $\times 3 \times 3$ }\\
~  & ~ &\multicolumn{4}{c}{\centering  $M = 10^{6}M_{\odot}, a = 0.0$, $z$ enumerates $0.1, 0.2, 0.3$} &~ \\
~  & ~ &\multicolumn{4}{c}{\centering  $M = 10^{6}M_{\odot}, a = 0.98$, $z$ enumerates $0.1, 0.2, 0.3$} &~\\
\end{tabular}
\end{center}
\end{ruledtabular}
\end{table*}

\subsection{Testing phase}

Once a \ac{CNN} is trained, we test the performance with 7 different groups of testing data, the properties of which are introduced in Table
\ref{tab:simulated_wave}.  The setup of these groups is motivated by the
testing of two main points, namely the validity of the \ac{CNN}, and the
sensitivity of the \ac{CNN}.  For testing the validity, we prepare three groups
of signal+noise samples: signals that follow the same distribution as the training data,  signals that follow an astrophysically motivated
distribution, and signals produced by \ac{AAK} but with the same distribution
as the training data.  For testing the sensitivity, we apply the \ac{CNN} on
top of the remaining four groups, in which many intrinsic parameters are fixed,
but either mass, redshift, or spin is enumerated over a certain list.
Taking group 4 as an example, it replaces the \ac{SNR} distribution in group 1 with
a set of fixed \acp{SNR}, taking values $10,\  20,\ 
30,\ \ldots,\ 130$. Similarly, in group 5, it enumerates the mass of the \ac{MBH},
taking values of $10^4,\ 10^{4.5},\ \ldots,\ 10^7\ M_{\odot}$. In group 6, it
enumerates the spin of the \ac{MBH} in a similar way, taking values of $0.0,\ 0.2,\ 0.4,\ 0.6,\ 0.8,\ 0.98$ with fixed \ac{MBH} mass is $10^6\ M_\odot$. In group 7, it focuses on \ac{EMRI} systems with three kinds of \acp{MBH}: 1. $M = 10^{5.5}\ M_{\odot}$ and $a = 0.98$; 2. $M = 10^6\ M_{\odot}$ and $a = 0.98$; 3.  $M = 10^6\ M_{\odot}$ and $a = 0 $. 
And they enumerates the redshift, taking the value of $0.1,\  0.15,\  0.2,\  0.25,\  0.3,\  0.35. $

The trained \ac{CNN} takes the testing data and outputs a probability $y_{\rm
pred}$, which predicts the probability that the data contains a signal.
Conversely, $1-y_{\rm pred}$ predicts the probability that it is purely
Gaussian noise.

\section{Results}{\label{sec:Results}}

\subsection{Validity}

To measure the effectiveness of the trained \ac{CNN} model, we used the testing data of groups 1-3  in Table \ref{tab:simulated_wave}.
One can compare the efficiencies among detection methods through the \ac{ROC} curve, where the \ac{TAP} is plotted against the \ac{FAP}. 

In the calculation of the \ac{ROC}, one first chooses a value of $y^*$ as a detection threshold.
Any value of $y_{\rm pred}>y^*$ is regarded as an alarm.
At a given $y^*$, the \ac{FAP} and the \ac{TAP} is constructed as the fraction of {\it noise-only} and {\t signal plus noise} samples, respectively, that are reported as an alarm.
By varying $y^*$, one obtain pairs of \ac{TAP}-\acp{FAP}, and a line that links all these pairs is the \ac{ROC} curve.
By definition, the \ac{TAP} is a monotonic function of the \ac{FAP}.  

\begin{figure}[htbp]
\centering
\includegraphics[scale=0.6]{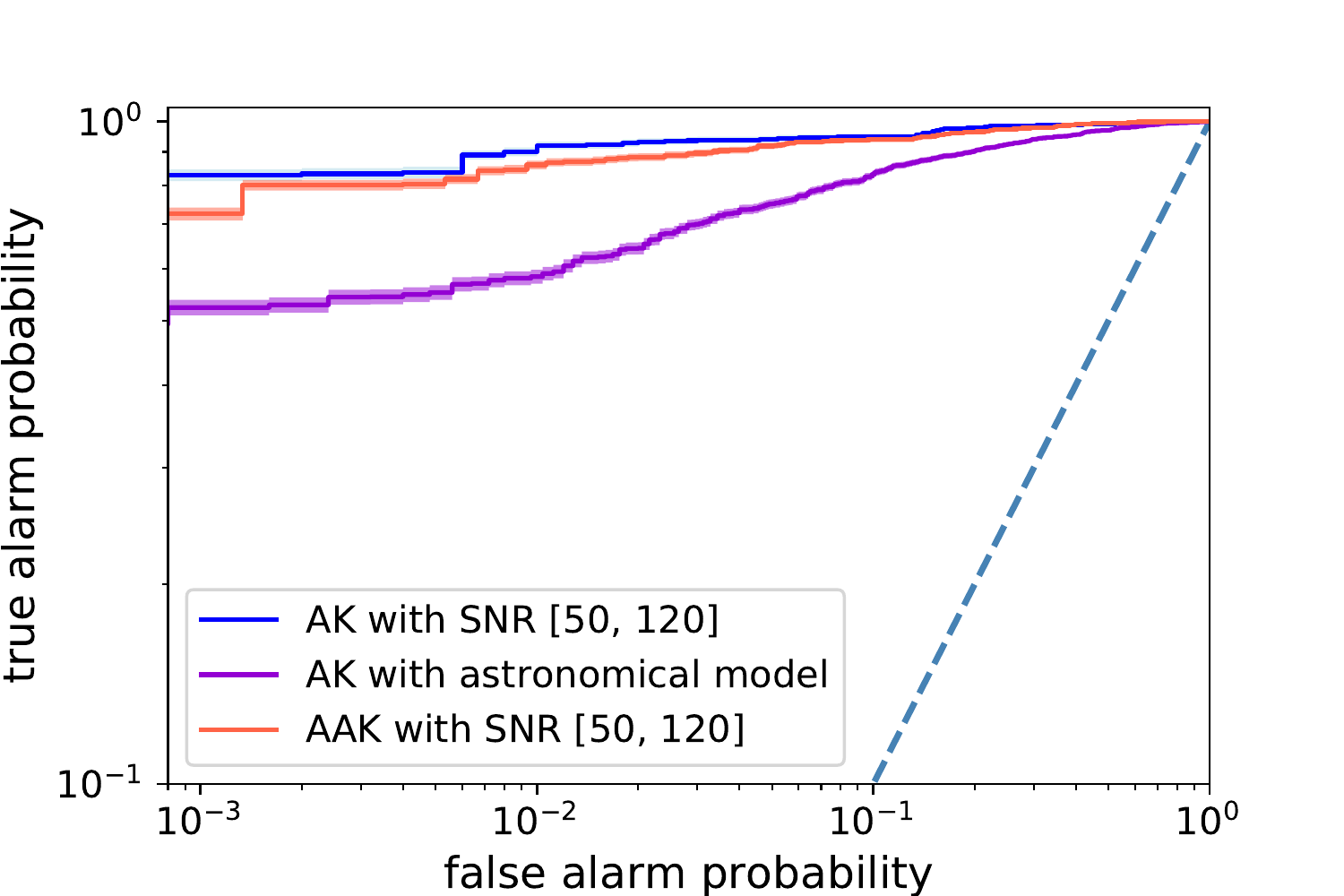}
  \caption{The \ac{ROC} curve of the signals from testing groups 1-3 is shown
with the blue, purple, and red lines, respectively. The blue line indicates
the expected effectiveness for group 1, the parameters have identical distribution
to the training data; for group 2, the distribution is drawn from an
astrophysical model; for group 3, the distribution is the same as group 1 and
the training data, but switched to the \ac{AAK} waveform model. The 1-$\sigma$ confidence intervals are indicated by the shaded regions.}
\label{fig:ROC}
\end{figure}

The blue line in Figure \ref{fig:ROC} represents a \ac{ROC} curve for group 1.  The dotted line is the worst possible performance which indicates a total
inability of distinguishing signal from noise.  One can observe that for
\ac{FAP} equal to 1\%, the corresponding \ac{TAP} is approximately 92\%.  This fact indicates that when the \ac{CNN} is applied to data constructed in a similar way to training data, it has a very good ability to distinguish the signal from noise.
We remark that the results from group 1 are expected to be the upper limit
among all groups and could be used as a benchmark.

The purple line in Figure \ref{fig:ROC} corresponds to the \ac{ROC} curve for
group 2.  Compared with the blue line, the purple line is lower: for a \ac{FAP}
of 1\%, the \ac{TAP} is about 75\%.  Since group 2 is an astrophysically
motivated population, the majority of signals are expected to be associated
with relatively low \acp{SNR}.  Due to the choice of \ac{SNR} threshold of 50 for the training data,
 we observe clustering of \acp{SNR} just above 50.  This significant
decrease in \ac{TAP} can be explained by such clustering.
On the other hand, even though the training data differ in the parameter
distribution to the testing data, the trained \ac{CNN} still has a good ability
to detect signals, showing a good generalization ability.

The red line in Figure \ref{fig:ROC}  corresponds to the \ac{ROC} curve of
group 3.  In this group, the parameter distribution is the same as in group 1,
but a different set of waveform models, namely the \ac{AAK} waveform, is
adopted.  This group could be used to study the robustness of the \ac{CNN}
against choices of \ac{EMRI} waveforms.  The similarity between the blue line
and the red line indicates that a \ac{CNN} trained on \ac{AK} waveforms can
achieve good detection ability on \ac{AAK} waveforms.  One should be cautious
not to take this conclusion for granted.  Although both \ac{AK} and \ac{AAK}
adopt similar equations, the setup of the waveform calculation makes them
dephase quickly after a timescale of around an hour.  This observation gives us
confidence that the \ac{CNN} method could be robust, and even if a fully
realistic waveform would not be available when space-borne \ac{GW} detectors
are operating, one can still rely on \ac{CNN} trained on approximate waveforms
to achieve good detection potential up to certain \ac{SNR} values.

\subsection{Sensitivity}

\begin{figure*}[t!]
	\centering
	\subfigure[\label{fig:efficiency-SNR}]{
    \includegraphics[scale=0.5]{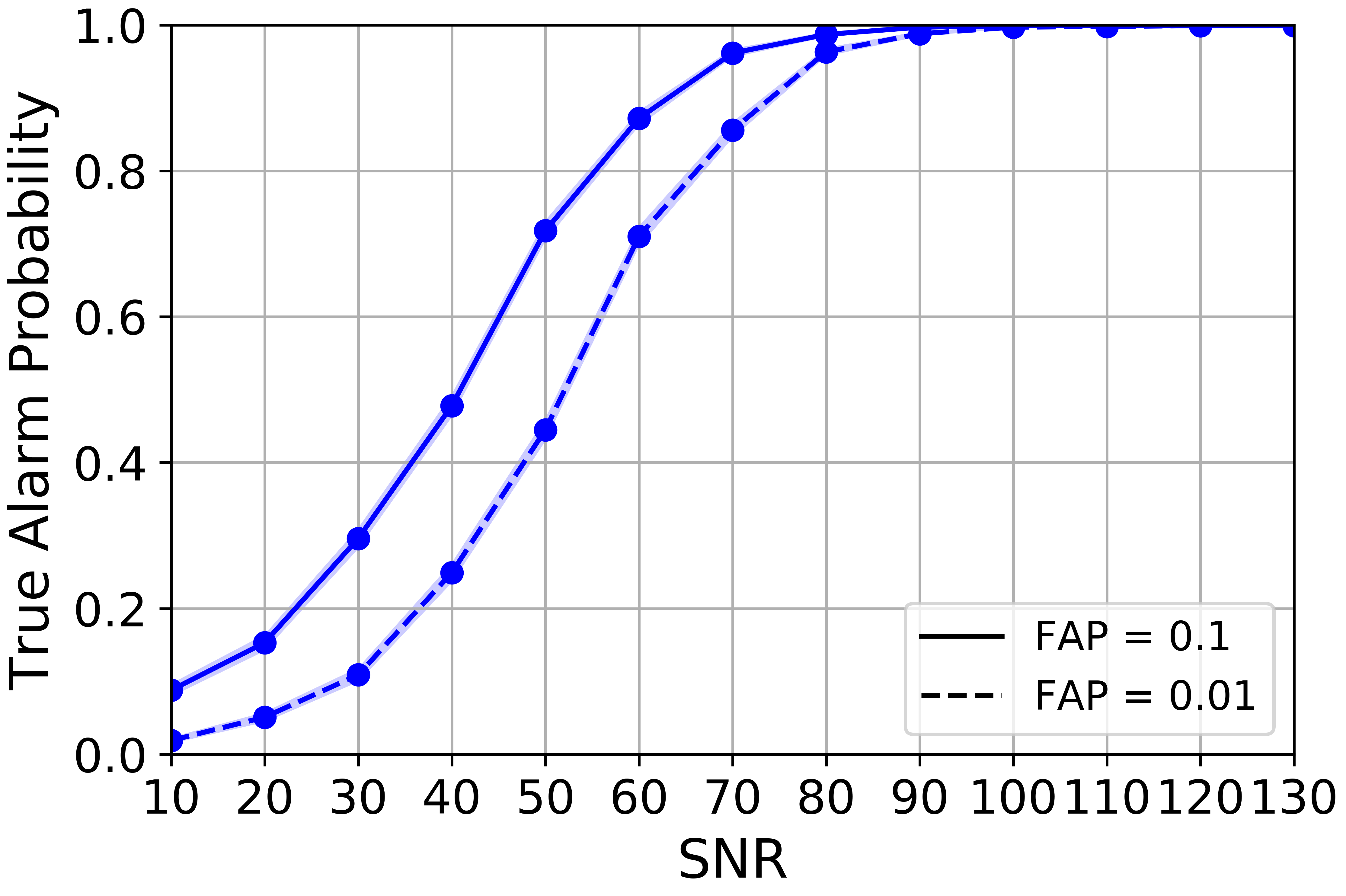}
    }
	\subfigure[\label{fig:efficiency-SMBHmass}]{
		\includegraphics[scale=0.5]{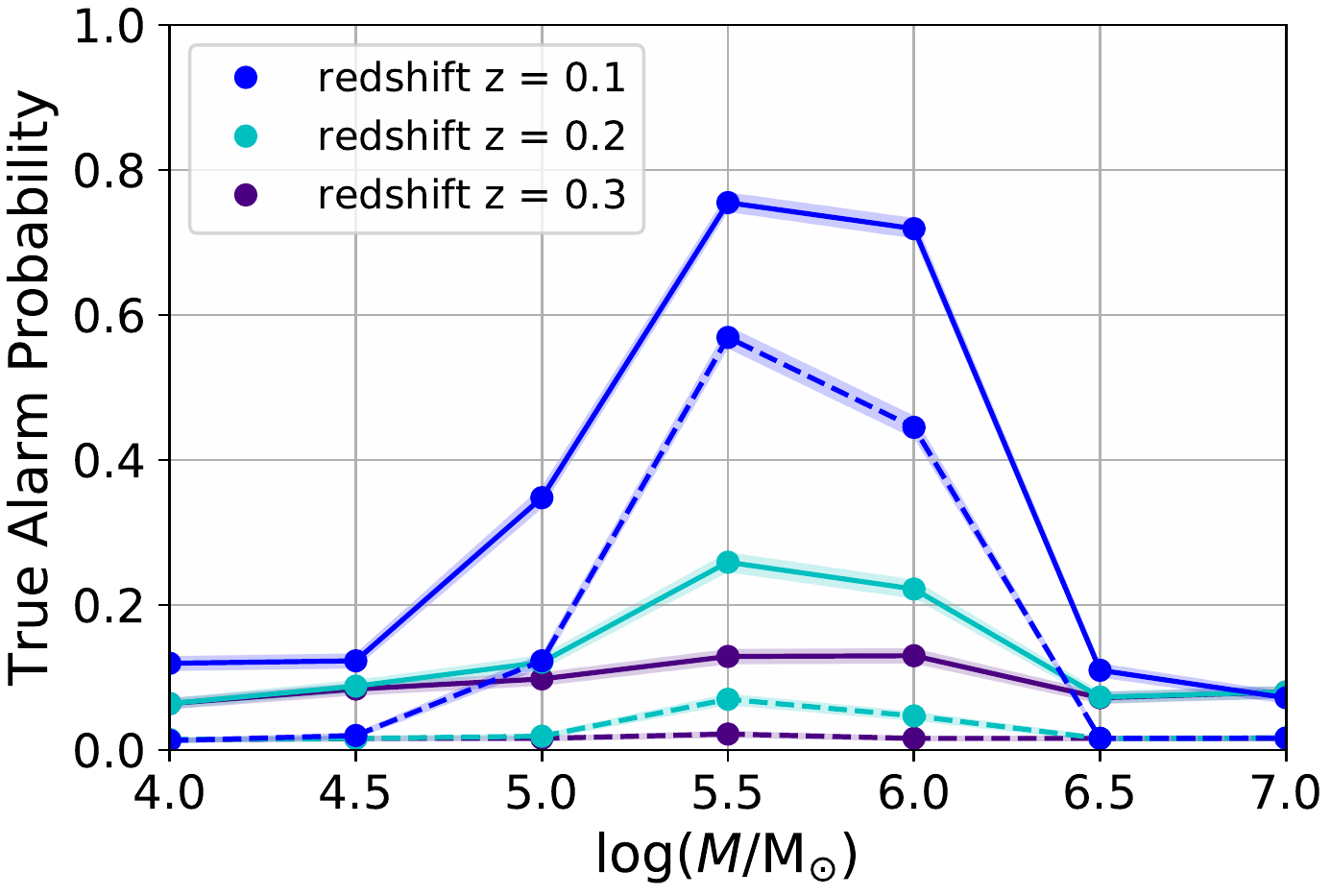}} \\
	\subfigure[\label{fig:efficiency-Spin}]{
		\includegraphics[scale=0.5]{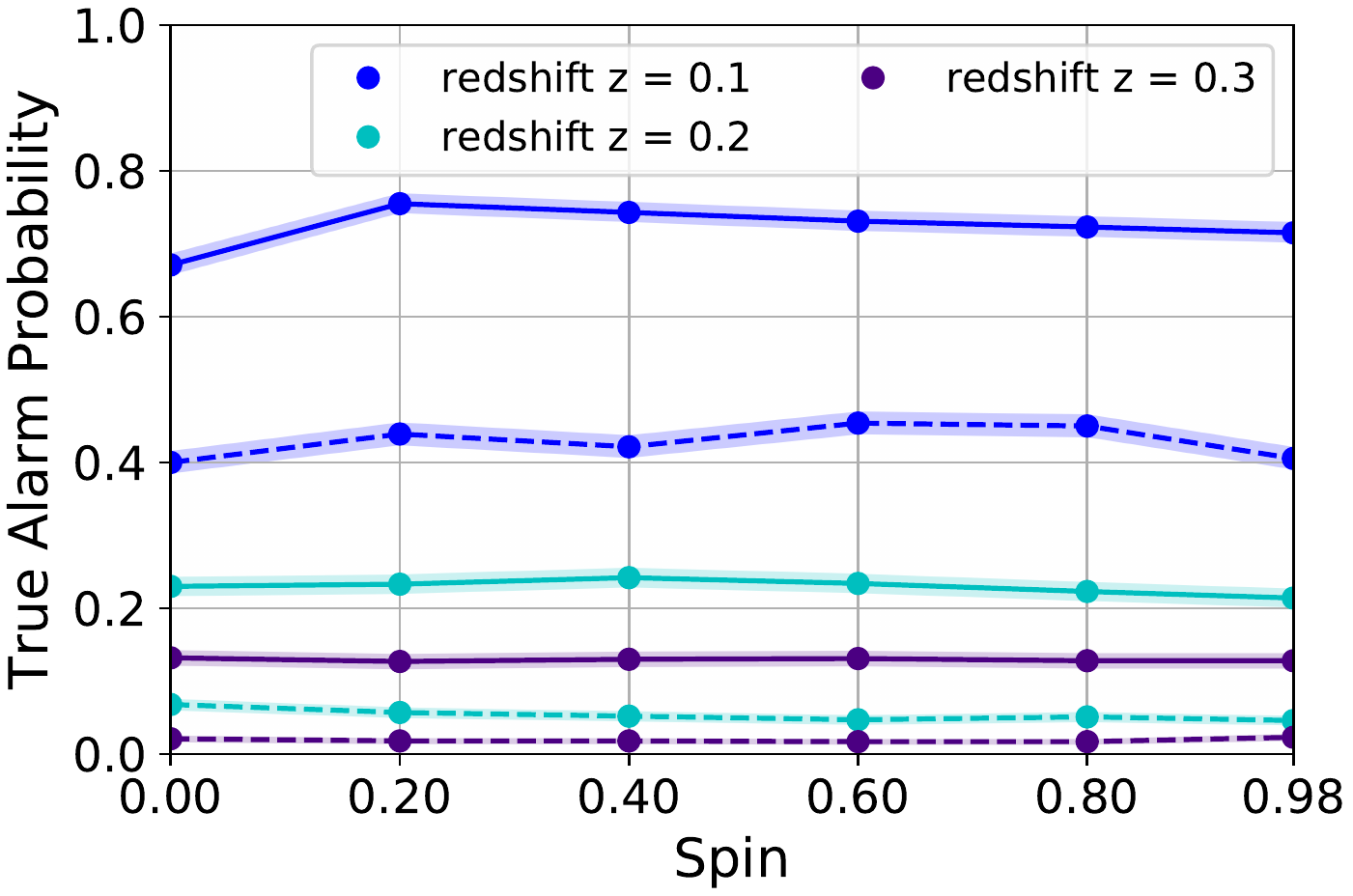}}
	\subfigure[\label{fig:efficiency-Redshift}]{
		\includegraphics[scale=0.5]{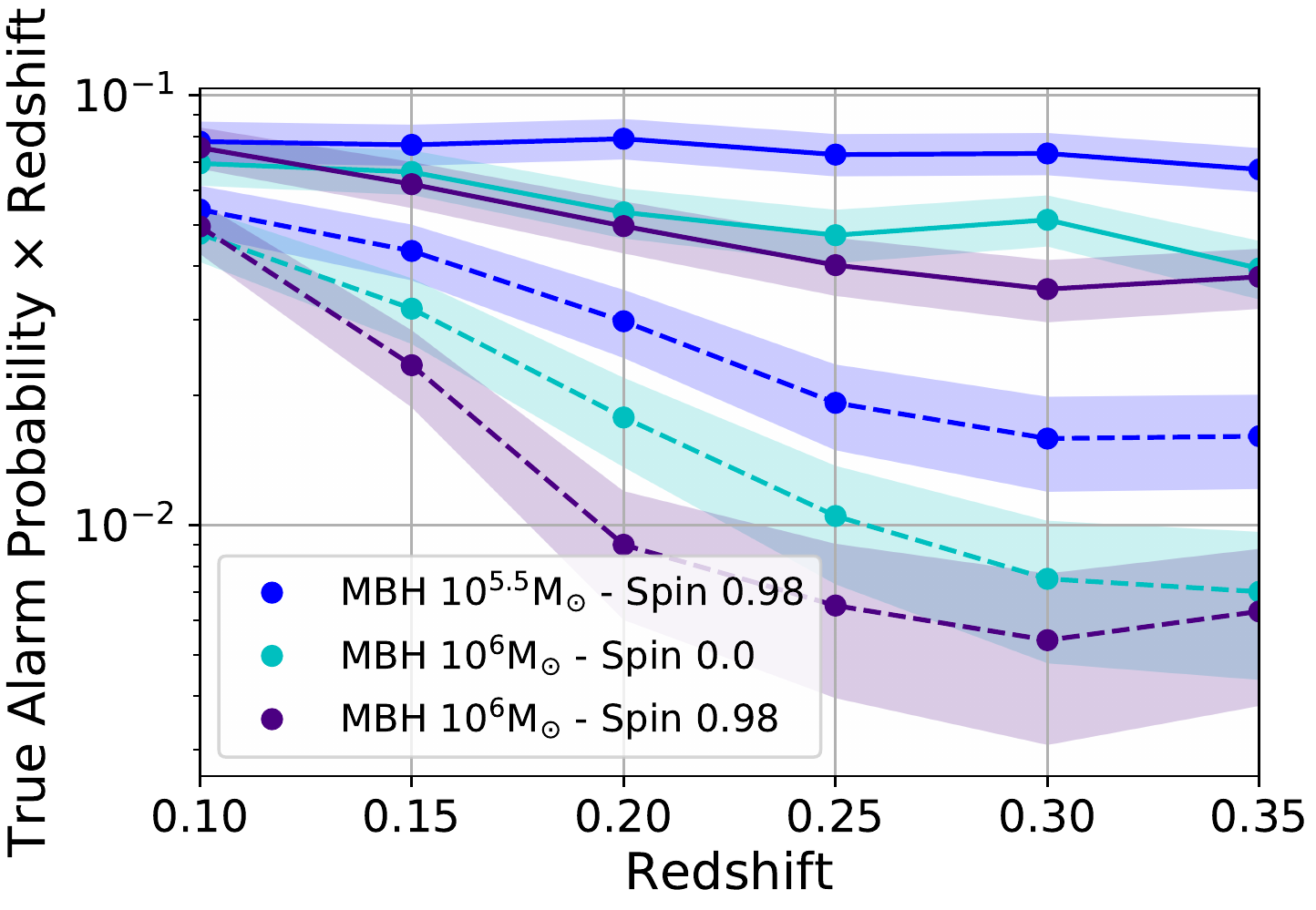} }
	\caption{The comparison of the \ac{CNN} sensitivity over EMRIs with different parameters.The vertical axis is the \ac{TAP}, while the horizontal axis is the single
varying parameter. The 1-$\sigma$ confidence intervals are indicated by the shaded regions.}
\label{fig:efficiency}
\end{figure*}

One can present the sensitivity of the \ac{CNN} over parameters like \ac{SNR}, \ac{MBH} mass, \ac{MBH} spin, or redshift, using the distribution of signal parameters group 4-7 in Table \ref{tab:simulated_wave}.
We have adopted fixed \acp{FAP} of 0.1 and 0.01.
The results are presented in Figure \ref{fig:efficiency} in the form of efficiency curves.

%
Figure \ref{fig:efficiency-SNR} displays the \ac{TAP} versus the \ac{SNR} for signals from group 4. 
It shows that two efficiency curves are consistent with the expectation that the \ac{CNN} exhibits higher sensitivity towards stronger signal,
and for \ac{SNR} of higher than about 100, the \ac{CNN} both can guarantee a certain detection at false alarm probabilities of 0.01 and 0.1.

We study the effectiveness of the \ac{CNN} over \ac{MBH} masses with group 5.
The \ac{MBH} spin is fixed as $a = 0.98$, and redshifts are varied through $z =
0.1,\  0.2,\  0.3$.  It can be seen from Figure \ref{fig:efficiency-SMBHmass} that
\ac{EMRI} systems containing a $10^{5.5} M_{\odot}$ \ac{MBH} are more
easily detected by the trained \ac{CNN}. For such an \ac{EMRI} with a redshift
of $0.1$, the \ac{CNN} can achieve a \ac{TAP} of 78\% conditioned on a
\ac{FAP} of 0.1. The feature of a sensitivity peak with regards to
\ac{MBH} masses stems from two factors: the sensitivity curve,
and the GW signal frequency distribution. Larger \ac{MBH} masses lead to larger amplitudes and lower
frequencies, while the detector is most sensitive only for a limited frequency
band. This result is also consistent with horizon distance analysis in a previous study~\cite{Fan:2020zhy}.


\ac{EMRI} signals contained in group 6 have a fixed \ac{MBH} mass ($M = 10^6 M_{\odot}$) and fixed redshifts ($z = 0.1,\ 0.2,\ 0.3$).  
From Figure \ref{fig:efficiency-Spin} one can observe that there is almost no apparent dependency of detection ability on \ac{MBH} spin.  
This is a piece of indirect evidence that the detection ability is most sensitive to the \ac{SNR}.  
Further investigation shows that \ac{EMRI} systems with different spins share similar \acp{SNR}.
This can be roughly explained as follows: the \ac{EMRI} systems have a smooth evolution of overall \ac{GW} amplitude before the final merger.
According to our setup, the merger is fixed at the end of the three months observation period.
Therefore, the spin parameter has a minor impact on the overall \ac{SNR}.

We present the efficiency curve for group 7 in Figure
\ref{fig:efficiency-Redshift}.  The \ac{SNR} of a signal is roughly
inversely proportional to luminosity distance, which for the nearby universe is approximately
proportional to redshift.  Therefore, on the vertical axis we plot the \ac{TAP}
multiplied by the redshift.  If the \ac{TAP} is linearly correlated with the \ac{SNR}, then
such multiplication should remain constant.  This is exactly what we observed
in the blue solid line ($M=10^{5.5}M_\odot$ and spin =0.98, \ac{FAP} equals
0.1).  The distribution of \ac{SNR} within this redshift range (0.1-0.35)
roughly spans  the range of 20-60, which shows good linearity between \ac{SNR}
and \ac{TAP} in Figure \ref{fig:efficiency-SNR}.  The prediction is violated to
the biggest degree for the purple dashed line ($M=10^6M_\odot$ and $a = 0.98$,
\ac{FAP} equals 0.01).  Closer events are gaining more detection efficiency, 
which can be expected from the non-linearity of the dashed line in the similar \ac{SNR} range in Figure
\ref{fig:efficiency-SNR}.

From the above analyses, we can conclude that among all factors, the \ac{SNR} has
the strongest influence on the expected detection efficiency.  Changes in other
parameters can also lead to a different performance in \ac{TAP}, but such
differences can be mostly explained by the different \acp{SNR}.

\section{conclusions and Future works}\label{sec:conclusion}

The \ac{GW} detection of \ac{EMRI} signals is challenging from multiple perspectives.
The accurate modeling of the \ac{EMRI} waveform is a challenging task for theorists.
At the same time a template bank with $\sim 10^{40}$ templates is needed, if one is to adopt the traditional matched filtering method \cite{gair2004event}. 
In this work, we demonstrate a proof-of-principle application of a \ac{CNN} on the detections of \acp{EMRI} signals.
Specifically, an \ac{EMRI} source with a signal-to-noise ratio uniformly drawn from 50 to 120 can be detected with a detection rate of 91\% at a search false alarm probability of 1\%. 
Meanwhile, the effectiveness of the \ac{CNN} varies for different physical parameters, however, most of the difference can be explained by the influence on \acp{SNR}.
\ac{EMRI} systems associated with higher \ac{SNR} can be expected to be detected with a higher chance.

Our work indicates that detecting \ac{EMRI} signals with \ac{CNN} is appealing, also for multiple reasons. 
For example, after being properly trained, the \ac{CNN} can be applied to three months' data, and finish the classification within one second.
On the other hand, \ac{CNN} shows a good generalization ability against a change of waveform models.
While training on the less physical \ac{AK} waveform, the network can successfully recognize signals injected with \ac{AAK} waveforms. 
This implies that the \ac{CNN} method has the potential to achieve actual detections even without accurately modeled waveforms.

It is important to benchmark the performance of the \ac{CNN} to other methods.
However, to the best of the authors' knowledge, there is no other study that uses physical \ac{EMRI} waveforms to search the signal and also provides a \ac{FAP} estimate \cite{wang2012extreme}.
For example, some studies use physical waveforms to search for the signal, but no \ac{FAP} is available \cite{gair2004event}.
On the other hand, other studies can provide \ac{FAP} estimates, but under a very different context, and the authors adopt phenomenological waveforms for the detection \cite{wang2012extreme}.


Although this proof-of-principle study is encouraging, we recognize that there are still lots of challenges to implement a reliable \ac{CNN} to detect \ac{EMRI} signals. 
For example, one needs to push the \ac{SNR} threshold to the values lower than 50.
This translates to a much larger amount of training data, the data generation, data storage, and training procedure would all become more hardware demanding.
We leave the issue for future studies, with the potential of applying more sophisticated methods or constructing a hierarchical search and examining strategy for the detection pipeline.

At present, there are many \ac{ML} algorithms that have wider applications in \ac{GW} data analysis. 
For different problems in the \ac{GW} field, developers use different methods or different architectures of the same method like \ac{CNN}.
Similar challenges might also apply to the \ac{EMRI} search, and we could borrow the existing wisdom in the future. 

\section{Acknowledgments}

The authors thank Hui-Min Fan,  Lianggui Zhu, En-Kun Li, Jianwei Mei, and
Micheal Williams for helpful discussions.  This work has been supported by
Guangdong Major Project of Basic and Applied Basic Research (Grant No.
2019B030302001), the Natural Science Foundation of China (Grants No. 12173104,
No. 11805286, and No. 11690022). CM is supported by the Science and Technology
Facilities Council grant ST/V005634/1. The authors would like to thank the
Gravitational-wave Excellence through Alliance Training (GrEAT) network for
facilitating this collaborative project.  The GrEAT network is funded by the
Science and Technology Facilities Council UK grant no. ST/R002770/1. NK would like to thank the support of the CNES fellowship. 
The authors acknowledge the uses of the calculating utilities of \textsf{numpy}
\cite{vanderWalt:2011bqk}, \textsf{scipy} \cite{Virtanen:2019joe}, and the
plotting utilities of \textsf{matplotlib} \cite{Hunter:2007ouj}.

\bibliography{ref}

\end{document}